# Identifying Topological Invariants of Non-Hermitian Systems via Domain-Adaptive Multimodal Model for Mathematics

Jiuchun Meng, [1,§] Lichao Sun, [2,§] Xiumei Wang,[2] Dandan Zhu,[3*] and Xingping Zhou [4*]

*[1] Portland Institute, Nanjing University of Posts and Telecommunications, Nanjing 210003, China*

*[2] College of Electronic and Optical Engineering, Nanjing University of Posts and Telecommunications, Nanjing 210003, China*

*[3] School of Computer Science and Technology East China Normal University Shanghai, China*

*[4] Institute of Quantum Information and Technology, Nanjing University of Posts and Telecommunications, Nanjing 210003, China*

*[*] Author to whom any correspondence should be addressed.*

*§ These authors contributed equally to this work.*

*ddzhu@mail.ecnu.edu.cn

*zxp@njupt.edu.cn

Identifying topological invariants are endowed with profound physical connotations in the fields of condensed matter physics. However, they are often limited by the failure of standard theorems and high computational costs. Traditional machine learning methods typically treat this problem as a black-box regression, which fails to learn the underlying mathematical structures of lattices. To this end, we propose a straightforward yet powerful multimodal model that fundamentally improves topological invariants discovery through two key methodological innovations. First, instead of feeding raw data into a standard network, our model introduces a dual-track alignment mechanism. This mechanism treats eigenvalues as context sequences and eigenvectors as matrix structures, enabling the model to naturally capture the interrelated algebraic and geometric properties of topological states. Second, to resolve the common problem where deep learning models make numerical mistakes in exact mathematical calculations, we use a tool-integrated reasoning paradigm. Within this paradigm, the neural network acts as a logical controller to calculate discrete topological indices. This design eliminates the uncertainty of deep learning, ensuring zero-error calculations from continuous physical data. We demonstrate that our model can accurately reconstruct complex three-dimensional generalized Brillouin zones and achieve 97% accuracy in calculating non-Bloch Chern numbers in long-range transition systems. By combining the flexible logic of multimodal artificial intelligence with the rigorous precision of symbolic computation, this work provides a reliable and highly versatile tool for exploring exotic topological phases.

## INTRODUCTION

Tracking and computing quantized topological invariants has become a cornerstone of condensed matter physics. For traditional Hermitian systems, Fukui et al. proposed a landmark approach in 2005 [1], constructing an explicit gauge-invariant formula using the overlap of discrete eigenvectors, thus efficiently computing Chern numbers on lattice grids. However, when extended to non-Hermitian systems, the traditional volume-boundary correspondence breaks down due to the non-Hermitian skin effect. To restore this fundamental correspondence, Wang et al. pioneered non-Bloch band theory by introducing the concept of generalized Brillouin zones (GBZs), which subsequently facilitated the reconstruction of non-Bloch Chern numbers. Recently, the mathematical elegance of this framework has been further extended to arbitrary high-dimensional spaces through the amoeba formula [2]. In parallel with these analytical and numerical developments, data-driven machine learning has emerged as an alternative paradigm, encompassing both supervised learning [3-13] and unsupervised learning [14-24] methods for identifying topological phases in a wide variety of physical systems.

Despite these theoretical successes, practical numerical computations remain highly challenging when moving into higher-dimensional systems. The main reason is that near a topological phase transition, the bulk gap closes, causing the Berry curvature to exhibit extremely sharp and localized divergences within the multi-dimensional Brillouin zone. To accurately capture these sharp singularities without missing critical transition points, the required sampling density increases exponentially with the number of dimensions, causing the computational cost to surge dramatically. Although adaptive mesh refinement schemes have managed to address this issue in certain four-dimensional systems to some extent, calculating topological invariants in even higher-dimensional spaces still triggers a severe curse of dimensionality. Consequently, efficiently resolving the precise geometric boundaries of both topological invariants and generalized Brillouin zones (GBZs) remains a major computational bottleneck.

To solve this problem, we establish a domain-adaptive multimodal model for mathematics to predict the GBZ. Different with prediction methods based on direct data-to-label mappings, we start from the underlying physics, formulate mathematical descriptions for eigenvalues and eigenvectors respectively. It can further capture their physical information and aligns the two representations. In this way, we achieve a unified description across different lattice systems. Additionally, we design a specific data processing pipeline to increase the information density. Then, our model is used to predict the GBZ of the Su-Schrieffer-Heeger (SSH) system with long-range hopping terms [25] and the two-dimension (2D) non-Hermitian lattice system [26]. To evaluate the performance of our model, we introduce three metrics: the domain similarity [27], the output consistency [28] and the accuracy of the Chern number calculation. Five available large language models (LLMs) are incorporated into our model for comparison and we find that Qwen2.5-Math significantly outperforms the other LLMs on the test data.

## Results

### 2D NON-HERMITIAN LATTICE SYSTEM

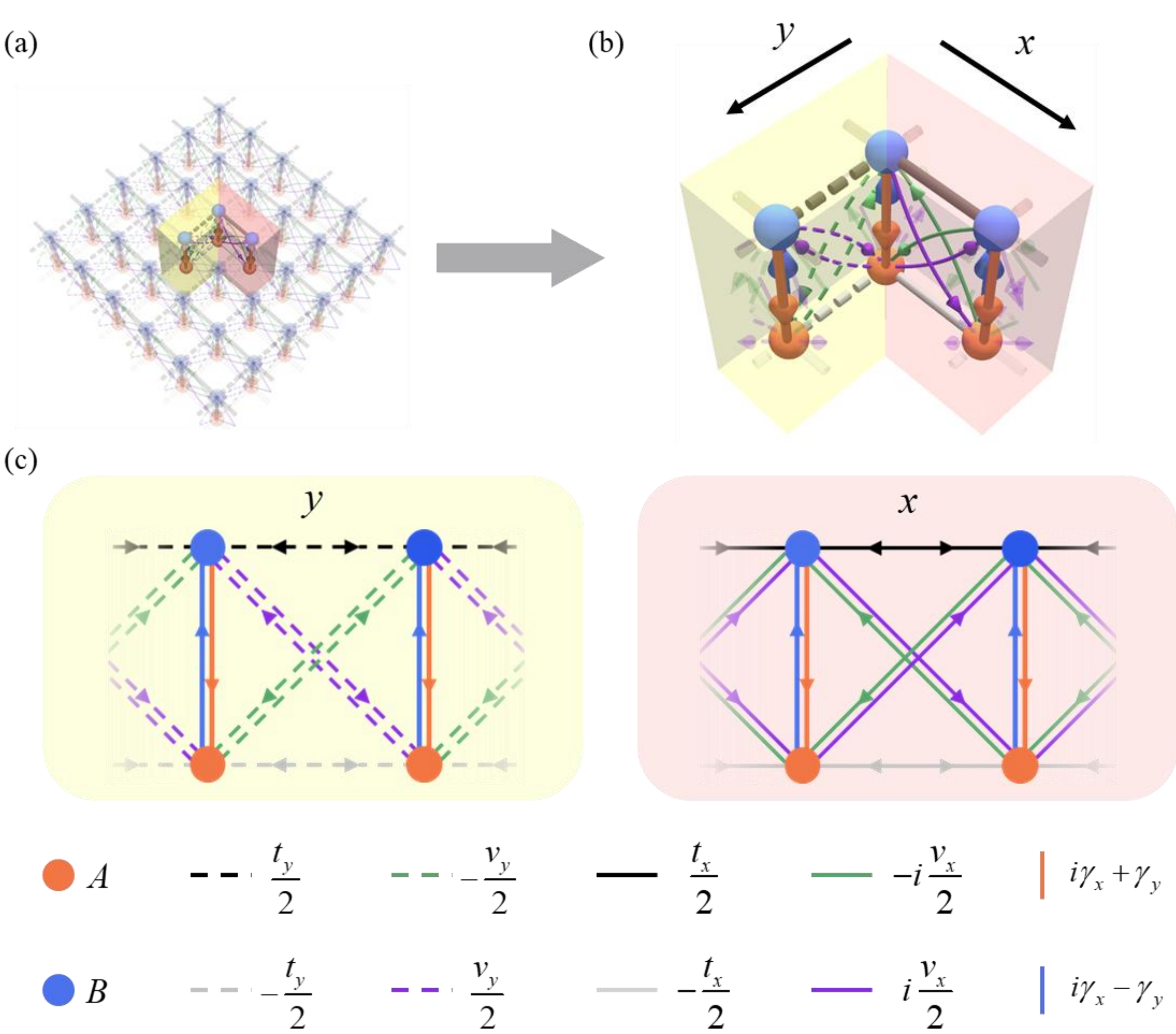

**FIG. 1. Illustration of the 2D non-Hermitian lattice system.**

The GBZ, a central concept in non-Bloch band theory, has provided a useful framework for describing non-Hermitian systems under the open-boundary conditions (OBCs), particularly for reproducing OBC spectra and formulating non-Bloch topological invariants [25]. However, the GBZ is generally difficult to determine explicitly, especially in higher-dimensional non-Hermitian systems. Here, we consider a 2D non-Hermitian lattice system [26], whose structure is schematically shown in Fig. 1. Since the non-Hermitian skin effect (NHSE) is suppressed in the $y$ direction of this lattice, the corresponding non-Bloch Hamiltonian is given by:

$$H(\beta_x, k_y) = \beta_x T_x + \beta_x^{-1} T_x^{\dagger} + e^{ik_y} T_y + e^{-\mathrm{i}k_y} T_y^{\dagger} + M, \tag{1}$$

where

$$T_x = \frac{1}{2}\begin{pmatrix} -t_x & -iv_x \\ -iv_x & t_x \end{pmatrix},\ T_y = \frac{1}{2}\begin{pmatrix} -t_y & -v_y \\ v_y & t_y \end{pmatrix},$$

$$M = \begin{pmatrix} m & i\gamma_x + \gamma_y \\ i\gamma_x - \gamma_y & m \end{pmatrix}, \tag{2}$$

As follows from Eqs. (1) and (2), the 2D non-Hermitian lattice consists of two sublattices $A$ and $B$, with onsite potentials $m$ and $-m$, respectively. As illustrated in Fig. 1, sublattices $A$ and $B$ are represented by the orange and blue sites, respectively. Here, couplings along the $x$ and $y$ directions are denoted by solid and dashed lines, while colors are used to distinguish different couplings. In particular, the vertical bonds within each unit cell are used to represent the intracell non-Hermitian couplings $i\gamma_x + \gamma_y$ on sublattice $A$ and $i\gamma_x - \gamma_y$ on sublattice $B$, respectively. By contrast, the same-sublattice couplings along the $x$ and $y$ directions are reciprocal. Along the $x$ direction, the corresponding hoppings are $-t_x/2$ for sublattice $A$ and $t_x/2$ for sublattice $B$; along the $y$ direction, they are $-t_y/2$ and $t_y/2$, respectively. Moreover, the couplings between different sublattices are given by $-iv_x/2$ and $iv_x/2$ for the $T_x$ term, and by $v_y/2$ and $-v_y/2$ for the $T_y$ term. For simplicity, all parameters are taken to be real.

The intricate geometry of the full GBZ makes it highly difficult to predict. Existing neural networks (e.g., convolutional neural networks) [3-4, 29] excel at small-output tasks like predicting topological invariants, but the dense, high-dimensional nature of the complete GBZ requires conventional learning strategies and data representations.

## DOMAIN-ADAPTIVE MULTIMODAL MODEL FOR MATHEMATICS

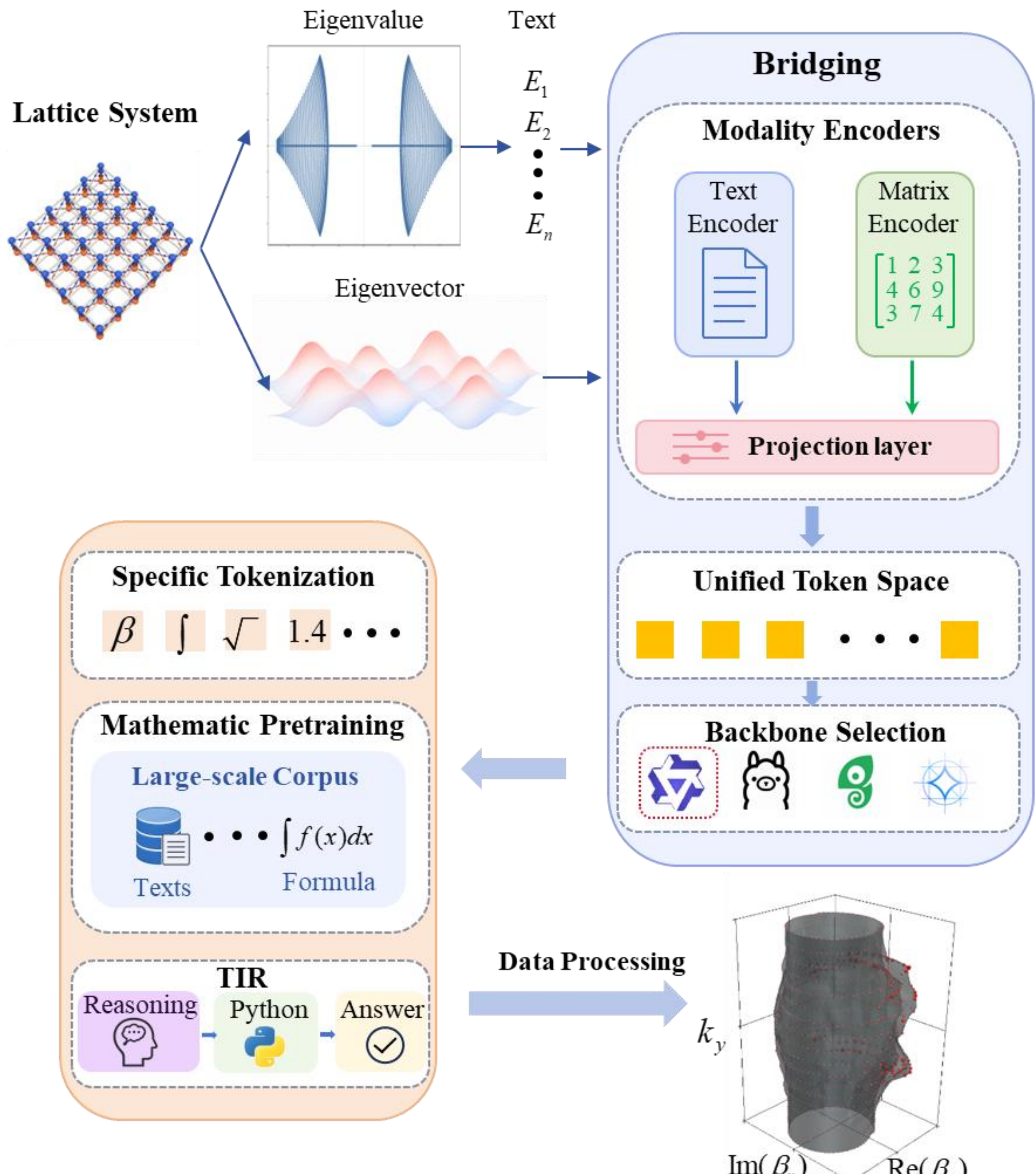

**Fig. 2: The architecture of domain-adaptive multimodal model for mathematics.**

Therefore, given the complexity of the lattices, we introduce the multimodal framework Align-Anything [30] to predict the GBZ of this complex lattice system. In prior works [3-4], the inputs were typically unimodal, with the data being represented in a single form (normally textual form). Such an input strategy is able to effectively accomplish the study of one specific lattice system. For example, some work sampled the Hamiltonian in momentum space and used it to characterize the physical properties of the lattice system [3]. Some other works sampled the complex energy spectrum as input for learning the winding number of the lattice system [4]. Inspired by these works, we consider making some improvements by introducing the eigenvalues and eigenvectors in momentum space as two input modalities, which may help fundamentally improves topological invariants discovery.

Conventional LLMs like GPT-4 [31], Qwen3 [32] and DeepSeek [33], are primarily designed for natural-language-centered tasks. Their performance is often limited when addressing specific and relatively complex mathematical problems. For instance, the widely used backbone Stable Diffusion [34] can capture certain visual features, but cannot effectively learn the potential physical logic of complex physical systems [35]. Moreover, its sensitivity to numerical precision is limited, and substantial errors often arise in fine details when complex images are characterized [35]. To reveal the impact of these deficiencies, we employ Stable Diffusion [34] to predict the GBZ curves for the one-dimensional (1D) non-Hermitian SSH model [25] and the SSH model with long-range hopping terms [25]. The predicted results are unsatisfactory (see the Supplementary Sec. I). It can be found that the shapes of the GBZs predicted by Stable Diffusion differ substantially from the standard GBZs.

To address this issue, we choose to employ the multimodal framework Align-Anything [30]. Align-Anything offers the fundamental functions of a multimodal LLM, enabling the reception and alignment of information from multiple modalities. In our research, Align-Anything plays the role of a physics-aware perceptual interface. At the beginning, our model introduces a dual-track alignment mechanism. In this mechanism, the eigenvalues and eigenvector are fed into the Align-Anything framework as context sequences and matrix structures respectively, to naturally capture the interrelated algebraic and geometric properties of topological states. Align-Anything then decodes the inputs individually, achieving both compression and alignment of high-density physical information. The compressed physical information will be mapped into a unified token space. In this space, all physical information is expressed as tokens that the LLM can readily understand and learn from.

In contrast to some closed end-to-end multimodal LLMs, Align-Anything is more akin to a plug-and-play multimodal framework. Align-Anything provides a natural interface for different backbones, thereby enabling a decoupled architecture between multimodal alignment at the front end and LLM reasoning at the back end [30]. The front end receives, encodes and aligns physics-derived information from different sources, whereas the back end can be flexibly coupled with different backbone to perform reasoning in the unified token space. In other words, Align-Anything provides a perceptual pathway from physical observables to our model, while the capacity for mathematical understanding and reasoning is primarily determined by the selected backbone. This design can be understood by analogy to the relationship between the neural sensory system and the brain in the human body. The neural sensory system receives heterogeneous signals from different sources, and the brain integrates these signals and performs reasoning. Therefore, the choice of an appropriate backbone is crucial for improving the model's understanding and reasoning capability with respect to the GBZ calculations.

After the comparison, we incorporate Qwen2.5-Math [36] as the backbone, thereby creating a domain-adaptive multimodal model for mathematics. We select Qwen2.5-Math as the "reasoning brain" of our model, primarily owing to its specialized design for mathematical problems. Firstly, Qwen2.5-Math is pretrained on large-scale mathematical corpus, which enables it to acquire a fundamental understanding of mathematical language and symbols before performing reasoning [36]. In addition, a specialized tokenization, tailored to mathematical expressions, is adopted by Qwen2.5-Math [36]. This design enables our model to more effectively represent and interpret the formula symbols involved in the GBZ calculations. Qwen2.5-Math also incorporates the concept of Tool-Integrated Reasoning (TIR), integrating a Python interpreter directly into the reasoning process [36]. During the training, the neural network acts as a logical controller to calculate discrete topological indices, which helps eliminate the uncertainty of deep learning and ensure zero-error calculations from continuous physical data. With the aid of these designs, our model is able to better capture the numerical mapping relations underlying complex physical problems.

## GBZ DATA PROCESSING PIPELINE

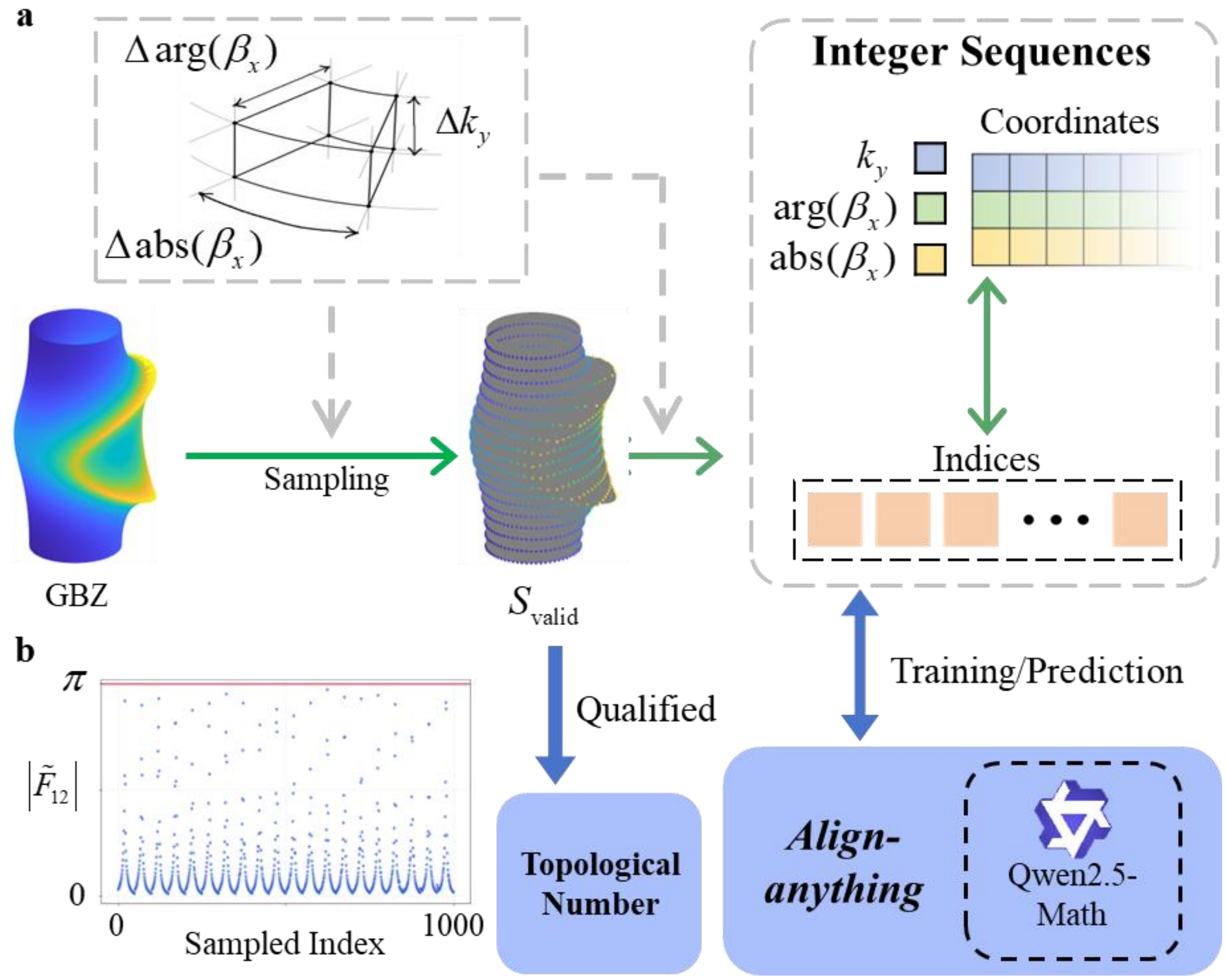

**FIG. 3. GBZ data processing pipeline.** (a) Numerically sampled GBZ coordinates $(k_y, \arg(\beta_x), \text{abs}(\beta_x))$ are encoded into discrete indices for training. During inference, the predicted indices are decoded to reconstruct the GBZ and evaluate the associated topological numbers. **(b)** The maximum value of $|\tilde{F}_{12}|$ for the 1000 samples of the data set.

Although our model possesses the ability to learn mathematical data, it remains a major challenge to effectively feed highly complex GBZ data into the model in a way that yields better learning performance. Recently, the amoeba-based formulation has been developed as a general framework for characterizing the NHSE and non-Bloch band theory in arbitrary spatial dimensions [2]. It provides a general formulation based on the amoeba and the Ronkin function, instead of exploiting all the boundary-condition equations whose number is proportional to the linear size in 2D OBC systems [2]. Another important advantage of this approach is that it yields a geometry-independent universal spectrum, or "amoebic spectrum", to which the OBC spectrum under any generic geometry converges [2]. As the authors pointed out, it is challenging to obtain a closed form for the Ronkin function in the 2D and higher dimensions [2]. In order to determine the GBZ of the higher-dimensional systems, one typically relies on numerical calculations of the minimum of the Ronkin function in practice. Therefore, the GBZ generally cannot be obtained analytically, and instead has to be determined numerically in the form of a discrete set of samples. At least for the Chern number, this discretization strategy is supported by the Fukui-Hatsugai-Suzuki (FHS) method [1], which establishes a gauge-invariant formulation on a discretized Brillouin zone.

Therefore, we discretize the GBZ into a three-dimensional (3D) orthogonal coordinate grid spanned by $k_y$, $\arg(\beta)$, $\mathrm{abs}(\beta_x)$, containing $N_{k_y}$, $N_{\arg(\beta_x)}$, $N_{abs(\beta_x)}$ grid points along the respective direction in our work. The corresponding grid spacings are defined as follows:

$$\Delta k_y = \frac{k_{y,\max} - k_{y,\min}}{N_{k_y} - 1},$$

$$\Delta \arg(\beta_x) = \frac{\arg(\beta_x)_{\max} - \arg(\beta_x)_{\min}}{N_{\arg(\beta_x)} - 1},$$

$$\Delta abs(\beta_x) = \frac{\mathrm{abs}(\beta_x)_{\max} - \mathrm{abs}(\beta_x)_{\min}}{N_{\mathrm{abs}(\beta_x)} - 1}. \tag{3}$$

For the 2D non-Hermitian lattice, since NHSE in the $y$ direction is suppressed [26], the Bloch wave number $k_y$ is real and ranges from $-\pi$ to $\pi$. In the $x$ direction, the non-Bloch factor $\beta_x$ is taken with $\arg(\beta_x) \in [-\pi, \pi]$. Moreover, the numerical data show that all sampled points satisfy $0.1 < |\beta_x| < 10$, and we set $|\beta_x|_{\min} = 0.1$ and $|\beta_x|_{\max} = 10$. The GBZ is then discretized using $N_{k_y} = 40$, $N_{\arg(\beta_x)} = 50$, $N_{abs(\beta_x)} = 120$. In practice, all data points are found to lie within this range.

By scanning all values of $k_y$ and $\arg(\beta)$, we obtain the full set of GBZ points, denoted by $S_{\mathrm{valid}}$ (see the Supplementary Sec. II). However, as a 3D set of real-valued points, $S_{\mathrm{valid}}$ contains a large number of non-terminating, non-repeating decimals. Since the LLM is constrained by a limited token budget, it is not well suited to directly process a massive amount of high-precision decimal data. To address it, we exploit the regularity of the sampling grid and convert the real-valued point set into a sequence of integers:

$$I = \left[\frac{\mathrm{abs}(\beta_x) - \mathrm{abs}(\beta_x)_{\min}}{\Delta abs(\beta_x)}\right] + \left[\frac{\arg(\beta_x) - \arg(\beta_x)_{\min}}{\Delta \arg(\beta_x)}\right] \times N_{abs(\beta_x)} + \left[\frac{k_y - k_{y,\min}}{\Delta k_y}\right] \times N_{\arg(\beta_x)} \times N_{abs(\beta_x)}, \tag{4}$$

where $[x]$ denotes rounding $x$ to the nearest integer. It should be emphasized that, in our GBZ calculations, the values of $k_y$, and $\arg(\beta)$ are obtained by exhaustively traversing all points on the predefined grid. Therefore, the terms associated with $k_y$ and $\arg(\beta)$ in Eq. (4) are necessarily integers, and only the $\mathrm{abs}(\beta_x)$ coordinates incur a finite loss of precision. This data-processing strategy can be generalized to GBZs of lattice systems in arbitrary dimensions, requiring only a modification of the dimensionality of the sampling grid.

In this way, the resulting integers represent coordinates on a regularly discretized complex space and naturally guarantee a strict inverse mapping to the original physical parameter space. Since in the present work the GBZ is represented by discrete samples, we compute the topological invariant using FHS [1]. In this approach, the lattice field strength $\left|\tilde{\boldsymbol{F}}_{12}\right|$ on the discretized points is required to satisfy

$$\left|\tilde{\boldsymbol{F}}_{12}\right| < \boldsymbol{\pi} . \tag{C1}$$

For the data set, we therefore evaluate the maximum value of $\left|\tilde{\boldsymbol{F}}_{12}\right|$ for each sample. As shown in Fig. 3(b), the data is valid. We also explored conventional neural network architectures, though these approaches yielded suboptimal results (detailed in Supplementary Sec. III).

## OUTCOMES AND EVALUATIONS

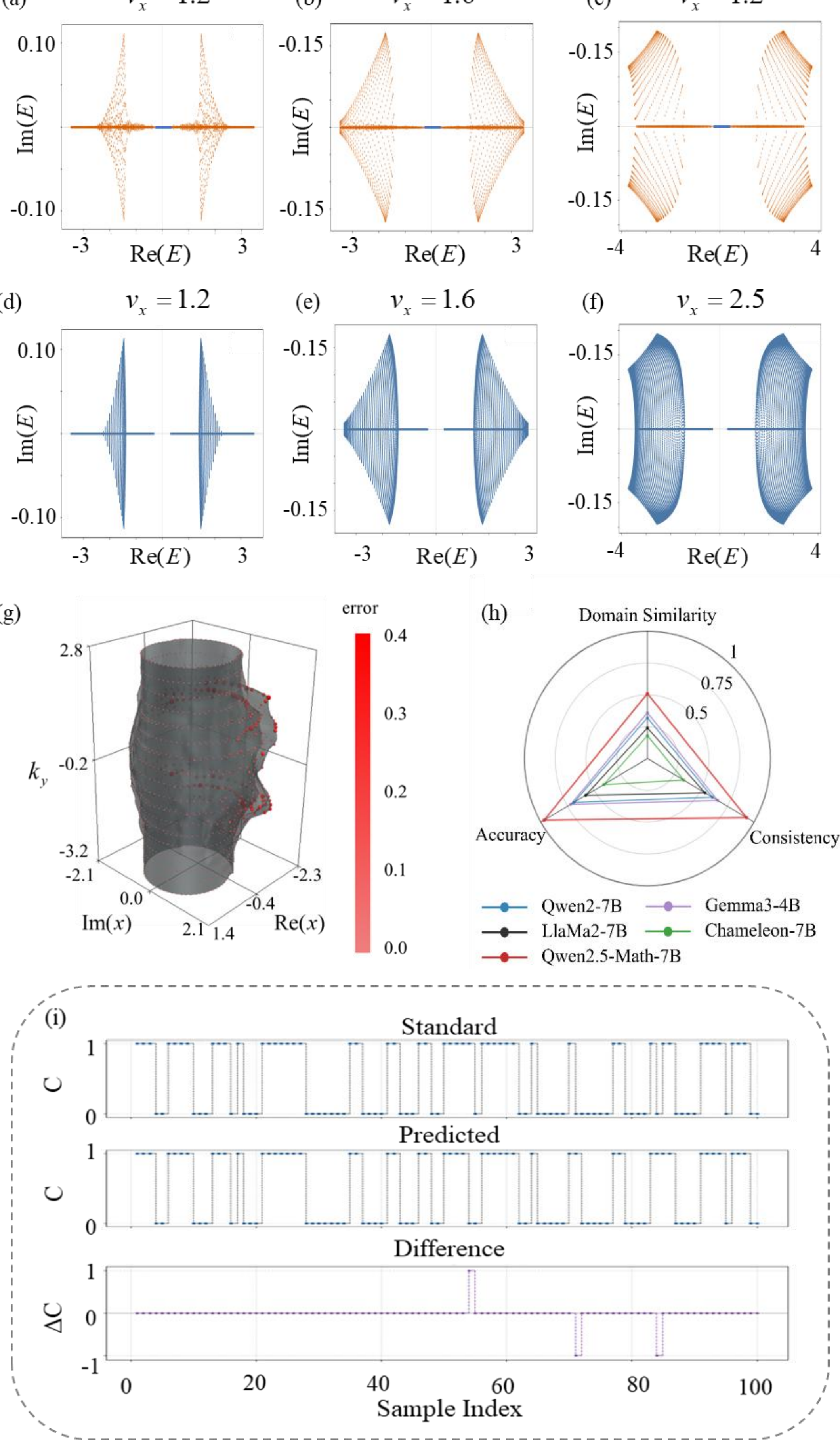

**Fig. 4.** (a-c) Energy bands calculated from the generalized Brillouin zone predicted by our model. (d-f) Energy level in the system with $L_x = L_y = 40$. We show the topological edge states in blue. (g) The predicted GBZ of 2D non-Hermitian lattice system by domain-adaptive multimodal model for mathematics. (h) The capability-evaluation radar chart for the five available LLMs. (i) Comparison between the predicted and standard Chern numbers for the 2D non-Hermitian lattice system. The parameters are taken as $t_x = 0.9$, $t_y = 1.0$, $v_x = 1.2$, $v_y = 0.8$, $\gamma_x \in [0.1, 0.2]$, $\gamma_y = 0$, $m \in [1.5, 2.5]$.

Based on the workflow outlined above, we construct a dataset containing 10000 GBZ samples of the lattice system for model training. We then adopt the LoRA finetuning strategy [37], in which the original information of the backbone model is kept frozen. Then, the learned weights are merged with the original information. This strategy greatly reduces the GPU-memory requirement of the computational process. Without this strategy, our hardware would be incapable of supporting the operation of our model. The final training results are stored in the form of model safetensors. To investigate whether our model has successfully captured the underlying mathematical relations, we employed it to predict the GBZ of the lattice under three distinct parameter configurations, and subsequently derived their corresponding energy spectra (Figs. 4a–c). For comparison, we also calculated the energy levels of the system on a finite open plane, as depicted in Figs. 4d–g. With the exception of the variable $v_x$, all other parameters were held constant: $t_x = 0.9$, $t_y = 1.0$, $v_y = 0.8$, $g_x = 0.2$, $g_y = 0.0$, and $m = 1.6$. A direct comparison of these spectra confirms the high accuracy of the GBZ predicted by our approach.

To further evaluate the effectiveness of our approach, we employed both the traditional method and our domain-adaptive multimodal large language model to calculate the GBZ under the parameter configuration: $t_x = 0.9$, $t_y = 1.0$, $v_x = 1.2$ $v_y = 0.8$, $g_x = 0.411$, $g_y = 0.0$, and $m = 1.949$. As shown in Fig. 4g, the gray surface represents the GBZ calculated by the traditional method, while the red points denote the GBZ generated by our model. Next, to quantify the computational accuracy, we compute the minimum distance to the reference GBZ surface for each point. As shown in Fig. 4g, a larger distance from the reference GBZ surface is indicated by a more intense color of the scattered point. We then compute the mean absolute distance from the entire point set to the reference GBZ surface, as defined by the following

formula:

$$D_i = \min_{S\in\text{GBZ}} \left( \min_{x\in S} \|p_i - x\|^2 \right),$$

$$D_{avg} = \frac{1}{N}\sum_{i=1}^{N} D_i \ , \tag{5}$$

where the $S$ refers to the partial surface of GBZ, and the $x$ refers to the random point on the surface $S$. The $p_i$ refers to one of the points in our output. The $D_i$ and $D_{avg}$ refer to the distance from point $p_i$ to the GBZ and the mean absolute distance respectively. Our calculation yields a mean absolute distance of 0.000466 for the GBZ point set obtained in our work. This result demonstrates that, the GBZ produced by our model agrees closely with the reference surface. We also use our model to predict the GBZ and calculate the winding number of the 1D non-Hermitian SSH model and the SSH model with long-range hopping terms (see the Supplementary Sec. IV), and the outputs are accurate. It indicates that our model has effectively learned how to predict the GBZ.

To further assess the performance of our model, we introduce two metrics: the domain similarity [27] and the output consistency [28]. Domain similarity characterizes the extent to which an LLM is aligned with the statistical distribution of a target domain. A higher domain similarity indicates that the backbone of our model is better adapted to the GBZ computation task. In our work, it takes the following form:

$$DS = \frac{|V_{LLM} \cap V_{GBZ}|}{|V_{GBZ}|} \times 100\%, \tag{6}$$

where the $V_{LLM}$ refers to the pretraining domain of the LLM, and the $V_{GBZ}$ refers to the domain of GBZ calculation. For the output consistency, this metric characterizes whether the model can stably generate highly similar results when the same problem is sampled multiple times. The outputs of LLM would tend to be stable when it has clearly learned how to solve the problem. In our work, we evaluate the output consistency of GBZ under the same coupling coefficients, as given by the following formula:

$$Consistency = \frac{n_{\max}}{n} \times 100\%, \tag{7}$$

where the $n_{\max}$ refers to the number of the most frequent answers. The $n$ refers to the number of prompts, which is set to 100 in our work. To clearly compare the quality of the accessible backbones, we select four additional LLMs for comparison: Qwen2 [38], Gemma3 [39], LlaMa2 [40] and Chameleon [41]. The corresponding results are shown in Fig. 4(b), with the detailed data provided in the Supplementary Sec. V. According to Fig. 4(b), it can be found that our model exhibits a high degree of domain similarity and output consistency after incorporating Qwen2.5-Math. It suggests that our model has developed a deep understanding GBZ.

For a further physical assessment of our model, we introduce the metric of Chern number calculation accuracy. One of the primary applications of the GBZ is the calculation of topological invariants, which are used to characterize the topological properties of the lattice system. For the 2D non-Hermitian lattice system, the Chern number can serve to distinguish topological phases and identify topological phase transitions. Therefore, it is necessary to evaluate the accuracy of the Chern numbers calculated from the GBZs predicted by our model. We generate 100 GBZ samples for the lattice system with different coupling coefficients, and compute the Chern numbers defined from the GBZs. Let $|u_{R,n}(\boldsymbol{k})\rangle$ and $\langle u_{L,n}(\boldsymbol{k})|$ denote the right and left eigenvectors of Eq. (1) respectively, where $\boldsymbol{k}$ is the complex Bloch wave vector with $k_x$ defined by $e^{ik_x} \equiv \beta_x$. Here, $n$ expresses a band index, and we set $n =+$ and $n =-$ to represent the energy bands with $Re(E) > 0$ and $Re(E) < 0$, respectively [26]. The right and left eigenvectors satisfy $\langle u_{L,m}(\boldsymbol{k}) | u_{R,n}(\boldsymbol{k})\rangle = \delta_{m,n}$. Then, we define the Chern number as [26]

$$C_n = \frac{1}{2\pi i} \int_{T_\beta} d\boldsymbol{k} \mathrm{B}_n(\boldsymbol{k}), \tag{8}$$

$$\mathrm{B}_n(\boldsymbol{k}) = \left\langle \frac{\partial u_{L,n}(\boldsymbol{k})}{\partial k_x} \middle| \frac{\partial u_{R,n}(\boldsymbol{k})}{\partial k_y} \right\rangle - \left\langle \frac{\partial u_{L,n}(\boldsymbol{k})}{\partial k_y} \middle| \frac{\partial u_{R,n}(\boldsymbol{k})}{\partial k_x} \right\rangle, \tag{9}$$

where $\mathrm{B}_n(\boldsymbol{k})$ is the Berry curvature. The definition of Eq. (8) means that the Chern number is defined as the integral of the Berry curvature over the GBZ denoted by $T_\beta$. We calculate the Chern number with $n =-$ by FHS method [1]. The resulting Chern numbers are then compared with the reference values to evaluate the accuracy of the Chern number calculation. The Chern number calculation accuracy of our model is 97%. The results are shown in Fig. 4(b), and the detailed data are provided in the Fig. 4(i). In the Fig. 4(b), it can be seen that the radar profile of Qwen2.5-Math almost encloses those of the other four LLM, which strongly supports the rationality of adopting Qwen2.5-Math as the backbone of our model.

## Discussion

In this work, we focus on the calculation of the Chern number for the 2D non-Hermitian lattice system. A domain-adaptive multimodal model for mathematics is established by integrating Qwen2.5-Math into the multimodal framework Align-Anything to stably predict the complex 3D GBZs of the lattice system. We also introduce a dual-track alignment mechanism, which regards the eigenvalues and eigenvectors as context sequences and matrix structures respectively. For a better performance of our model, we use the tool-integrated reasoning paradigm, which successfully helps eliminate the uncertainty of deep learning. With the aid of these methods, our model accurately predicts the GBZ images of the lattice system. In addition, we introduce three metrics to assess our model: the domain similarity, the output consistency and the accuracy of the Chern-number calculation. Furthermore, we compare the performance of different LLMs within our framework, in order to discuss what kinds of models are better suited to GBZ calculations.

Overall, our study is built upon previous works on the GBZ and topological invariants [1, 2, 25, 26, 42]. Existing studies have already established a relatively clear mathematical framework for the computation of the GBZ. For instance, the amoeba theory has successfully generalized the non-Bloch band theory and the GBZ to arbitrary dimensions from a mathematical perspective [2]. These works have relied on the existing mathematical formulas, which can be regarded as the prior knowledge. If one wishes to avoid dependence on such prior knowledge, symbolic regression [43-49] may provide a route to explore potential analytical solutions for the GBZ in higher dimensions. Unfortunately, symbolic regression still has some limitations with regard to achieving this goal. For one thing, the operator libraries currently used in symbolic regression are usually relatively simple, which makes it difficult to handle more complicated mathematical operators, such as left and right eigenvectors. For another, symbolic regression tends to favor the simplest relations, whereas the mathematical structures involved in the GBZ are often considerably more complex. To address these difficulties, one may consider improving symbolic regression's ability to represent complex operational relations through large-scale pretraining. In addition, suitable constraints may be introduced during training to guide the symbolic regression toward expected analytical forms.

## ACKNOWLEDGMENTS

The authors thank for the support by National Natural Science Foundation of China under (Grant 12404365 and 62377011).

## DATA AVAILABILITY

The codes that used in our work are openly available [50]. The data that support the findings of this article are not publicly available upon publication because it is not technically feasible and/or the cost of preparing, depositing, and hosting the data would be prohibitive within the terms of this research project. The data are available from the authors upon reasonable request.